\documentclass[10pt,showpacs,bibnotes,notitlepage,twocolumn,balancelastpage,preprintnumbers,amsmath,amssymb,final,prb]{revtex4}
\usepackage{graphicx}
\usepackage{color}

\begin{document}

\title{Exotic Haldane Superfluid Phase of Soft-Core Bosons in Optical Lattices}
\author{Jian-Ping Lv$^{1,2}$\footnote[1]{phys.lv@gmail.com}}
\author{Z. D. Wang$^{2}$\footnote[2]{zwang@hku.hk}}
\address{$^{1}$ Department of Physics, Anhui Normal University, Wuhu 241000, China; \\
$^{2}$  Department of Physics and Center of Theoretical and Computational Physics,
The University of Hong Kong, Pokfulam Road, Hong Kong, China
}
\date{\today}

\begin{abstract}
 We propose to realize an exotic Haldane superfluid (HSF) phase in an extended Bose-Hubbard model on the two-leg ladder (i.e., a two-species mixture of interacting bosons). %and to probe it in optical lattices.
  The proposal is confirmed by means of large-scale quantum Monte Carlo simulations, with a significant part of the ground-state phase diagram being revealed. Most remarkably, the newly discovered HSF phase
 features both superfluidity and the non-local topological Haldane order. The effects induced by varying the number of legs are furthermore explored. Our results shed light on how topological superfluid emerges in bosonic systems.
\end{abstract}
\pacs{74.20.Rp, 03.65.Vf, 03.75.Hh, 67.60.Bc}
\maketitle
%%%

\section{Introduction}

Searching for novel topological phase is at the frontier of condensed matter research~\cite{Qi2010}. Recently, apart from significant interests on topological insulators~\cite{Hasan2010,Qi2011} and topological metals including semimetals~\cite{FS-classification,ZhaoWangPRL},  topological superfluid (TSF) phase, as an exotic quantum phase, has also been attracting more and more
attention. The TSF is not only of fundamental importance but also has potential applications for topological quantum computing~\cite{Kitaev2006,Nayak2008}. A variety of schemes have been proposed to realize fermionic TSFs~\cite{Read2000, Kitaev2000, Qi2009, Sato2009, Cooper2009, Roy2008, Zhou2011, Liu2012, Chen2013, Cao2014, ZhaoWang-TSC}.
However, an idea to realize bosonic TSF,
which is rather distinct from the fermionic one, and how to probe it in a controllable way are still badly awaited.

An ideal experimental platform to quantum-simulate a nontrivial condensed matter model is an optical lattice loaded with cold atoms, which has successfully emulated Bose-Hubbard Hamiltonian and demonstrated superfluid-Mott insulator (SF-MI) transition~\cite{Greiner}. This experimental achievement has led to the renewed interest for Bose-Hubbard-like models~\cite{Fisher1989,
%}; for a review, see Ref.~\cite{
Bloch2008}.

\begin{figure}
%\vspace*{0cm} \hspace*{-0cm}
\begin{center}
%%\quad \vspace{4cm}  %% TEMPORARY UNTIL FILE IS THERE
\includegraphics[angle=0,width=8cm,height=3cm]{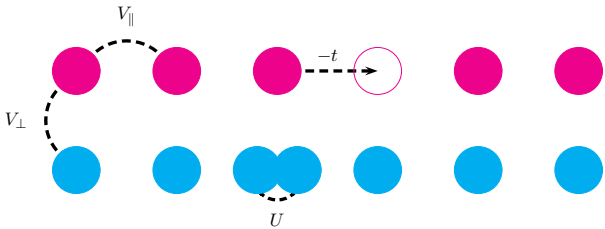}
\end{center}
\caption{(Color online) An illustration of present proposal [Hamiltonian~(\ref{hamiltonian})]. The bosons interact via a nearest-neighbor repulsion along chains ($V_{\parallel}$), a repulsion across chains ($V_{\bot}$), an onsite repulsion ($U$), and can hop to nearest-neighbor sites along chains with the energy $-t$. The bosons on two chains (of two species) are coloured in magenta and cyan, respectively.}
\label{guide}
\end{figure}

\begin{figure}
%\vspace*{0cm} \hspace*{-0cm}
\begin{center}
%%\quad \vspace{4cm}  %% TEMPORARY UNTIL FILE IS THERE
\includegraphics[angle=0,width=8cm,height=7cm]{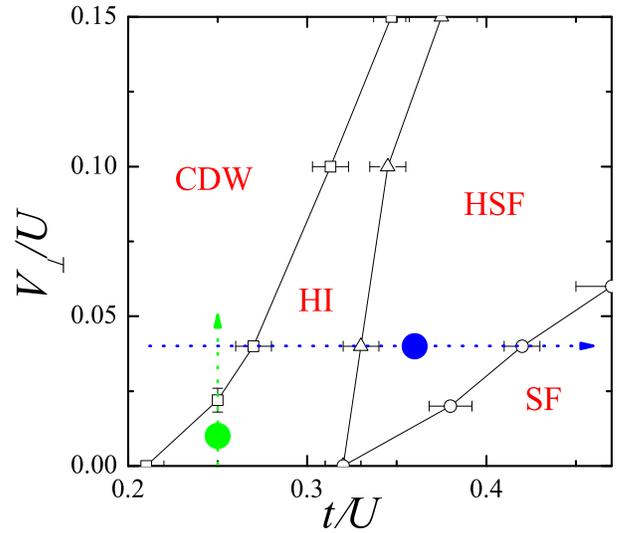}
\end{center}
\caption{(Color online) The ground-state phase diagram of Model~(\ref{hamiltonian}) with $V_{\parallel}/U=3/4$.
Quantum phases on the phase diagram include charge density wave (CDW), Haldane insulator (HI), Haldane superfluid (HSF),
 and superfluid (SF). The dotted green path is described in Fig.~\ref{t_0.25}, while the blue one is discussed in Figs.~\ref{V2_0.04}, \ref{scalings}, and \ref{merged}; the extensions to multi-leg ladders, corresponding to the two colored circles, are also illustrated in Fig.~\ref{merged}.}
\label{pd_v1_0.75}
\end{figure}

 It was recently indicated that Haldane insulator (HI) can appear in one-dimensional bosonic systems~\cite{Dalla Torre2006,Rossini2012,Batrouni2013,Batrouni2014,Ejima2014,Kurdestany2014}. The bosonic HI phase was initially found for dipolar bosons~\cite{Dalla Torre2006} and later in the extended Bose-Hubbard model which includes a nearest-neighbor boson-boson repulsion~\cite{Rossini2012,Batrouni2013,Ejima2014}. These HIs were determined in the case of unity filling,
where dominant occupation numbers are $0$, $1$ and $2$. The bosonic HI is therefore a reminiscent of the Haldane phase conjectured for spin-$1$ systems whose local spin variables are $-1$, $0$ and $1$~\cite{Haldane1983}. In a topological aspect, the Haldane phase is protected by the lattice inversion symmetry and can be classified as a symmetry-protected topological phase~\cite{Gu2009,Pollmann2012}.

A paradigmatic example of fermionic TSF is the inhomogeneous topological superfluidity which features modulated particle density and inhomogeneous SF on one-dimensional lattices~\cite{Zhai2015}.
Recall that the single-chain extended Bose-Hubbard model hosts a homogenous SF phase~\cite{Rossini2012,Batrouni2013,Batrouni2014,Ejima2014}.
When two such SF chains are coupled by an onsite repulsive interaction, the particle density modulation may arise along each chain.
The superfluidity thus becomes inhomogeneous and is possibly compatible with the non-local Haldane order.
As a result, a novel bosonic TSF phase, which we call Haldane superfluid (HSF) phase, may emerge.

The structure of the paper is as follows. Section~\ref{s2} introduces the model Hamiltonian.
Numerical method and physical observables are described in Sec.~\ref{s3}.
Section~\ref{s4} presents numerical results, and a brief discussion is given in Sec.~\ref{s5}.

\section{Model Hamiltonian}~\label{s2}
As a test bed for the aforementioned conjectures,
we consider an extended Hubbard model of soft-core bosons on two-leg ladder (shown in Fig.~\ref{guide}), which is described by
\begin{eqnarray}
\mathcal{H}=&-&t\! \sum_{\langle ij\rangle_{\parallel} } (a^{\dag}_i a_j +a_i a^{\dag}_j)+\frac{U}{2} \sum_{i} n_i (n_i-1)   \nonumber \\
  &+&  V_{\parallel} \sum_{\langle ij\rangle_{\parallel} }n_i  n_j+ V_{\perp} \sum_{\langle ij\rangle_{\bot} }n_i  n_j.
\label{hamiltonian}
\end{eqnarray}
Where $t$ represents the hopping amplitude along either chain of the ladder, $U$ is onsite repulsion,
$V_{\parallel}$ and $V_{\perp}$ are nearest neighbor repulsions respectively along and perpendicular to the chains.
Operator $a^{\dag}_i$ ($a^{ }_i$) denotes bosonic creation (annihilation) operator, and
$n_i=a^{\dag}_i a^{ }_i$ is particle-number operator. It is noteworthy that the system can be known as a
Bose-Bose mixture in the sense that each chain hosts a species~\cite{Boninsegni2001,Kuklov2003,Altman2003,Soyler2009,Lv2014}.

Here we focus on the case of unity filling where the Haldane string order can be stable.
The main findings are summarized as a ground-state phase diagram for $V_{\parallel}/U=3/4$, shown in Fig.~\ref{pd_v1_0.75}.
In the phase diagram we confirm the existence of topological phases, HSF and HI, in a broad parameter region.
In the following, we present Monte Carlo results to demonstrate the existence of HSF and HI phases and explore how they behave.

\section{Numerical method and physical observables}~\label{s3}
We perform large-scale Monte Carlo simulations for Model~(\ref{hamiltonian}),
using a unbiased algorithm of worm-type update that works in continuous imaginary time~\cite{Prokofev1998a,Prokofev1998b,Prokofev2009}.
 This algorithm is still efficient in rather difficult cases~\cite{lv2015}.
Our simulations are carried out in canonical ensemble with a broad range of chain length, $L \in [64,512]$.
To guarantee that the ground-state information is obtained, we decrease temperature until
its effect is negligible. The lowest temperature we use to check every conclusion is lower than $\beta =512$.

The following observables are sampled to reveal ground-state properties.
Superfluid density is evaluated via winding number fluctuations~\cite{Pollock1987}
\begin{equation}
\rho_{s}=  \frac{L^{2-d} \langle W^2 \rangle}{2\beta t},
\end{equation}
where $W$ represents winding number of a worldline configuration and $d=1$ is the effective dimension.
We also measure correlation functions that reveal crystalline order,  Haldane string order and parity,
which are respectively defined as:
\begin{eqnarray}
O_c(|i-j|)&=&(-1)^{|i-j|}<\delta n_i \delta n_j> , \nonumber \\
O_s(|i-j|)&=&<\delta n_i e^{i \theta \Sigma^{j-1}_{k=i} \delta n_k }\delta n_j>, \nonumber \\
O_p(|i-j|)&=&< e^{i \theta \Sigma^{j-1}_{k=i} \delta n_k }>.
\end{eqnarray}
Where $\delta n_i=n_i-\bar{n}$ represents the particle number fluctuation from average filling $\bar{n}$ (this work focuses on $\bar{n}=1$); $\theta=\pi$ denotes the topological angle, corresponding to the maximum of $O_s(\theta)$, in Haldane phase of spin-$1$ system that is analogous to the present model~\cite{Qin2003}. The measurements are taken along the chains, and the symbol $|i-j|$ denotes the distance between two lattice sites along the chains.
To explore long-range correlations, we pay special attention
to $O_c(L_{max})$, $O_s(L_{max})$ and $O_p(L_{max})$, with $L_{max}$ the maximum horizontal distance between two sites.
Our simulations are carried out on periodic lattices such that $L_{max}=L/2$.
Table~\ref{po} illustrates how these observables are used to distinguish among the quantum ordered phases.

\begin{table}\caption{The classification of quantum ordered phases --- CDW, SF, HI,
  MI, HSF and SS --- via the robustness of measured observables as
  $L \rightarrow \infty $ and $\beta \rightarrow \infty $.}\label{po}
\begin{tabular}{|c|c|c|c|c|}
  \hline

   & $\rho_{\rm s}$ &$O_c(L_{max})$ &$O_s(L_{max})$  & $O_p(L_{max})$ \\
      \hline
     CDW & $=0$ & $\neq 0$  & $\neq 0$  & $\neq 0$\\
     SF & $\neq 0$  & $=0$  & $=0$ & $=0$\\
     HI & $=0$ & $=0$  & $\neq 0$  & $=0$\\
     MI & $=0$ & $=0$ & $=0$   & $\neq 0$ \\
     HSF & $\neq 0$  & $=0$   & $\neq 0$  & $=0$ \\
     SS & $\neq 0$  & $\neq 0$  & $\neq 0$  & $\neq 0$\\
  \hline
\end{tabular}
\end{table}

\section{Monte Carlo results}~\label{s4}
We concentrate on a typical ratio of intrachain nearest-neighbour repulsion and onsite repulsion, the same as that in Ref.~\cite{Batrouni2013}, $V_{\parallel}/U = 3/4$. At $V_{\bot}=0$ Model~(\ref{hamiltonian}) reduces to the single-chain extended Bose-Hubbard model which hosts a bosonic Haldane insulating phase~\cite{Dalla Torre2006,Rossini2012,Batrouni2013,Batrouni2014,Ejima2014}. Our results consist with those in Ref.~\cite{Batrouni2013}: as $t$ increases there appear in order CDW, HI and SF phases, separated by a CDW-HI transition at $t/U=0.21(1)$ and a HI-SF transition at $t/U=0.32(1)$. In the following, we shall consider the cases that interchain repulsions are turned on, i.e., $V_{\bot} > 0$.

\begin{figure}
\centering
\includegraphics[angle=0,width=8cm,height=10cm]{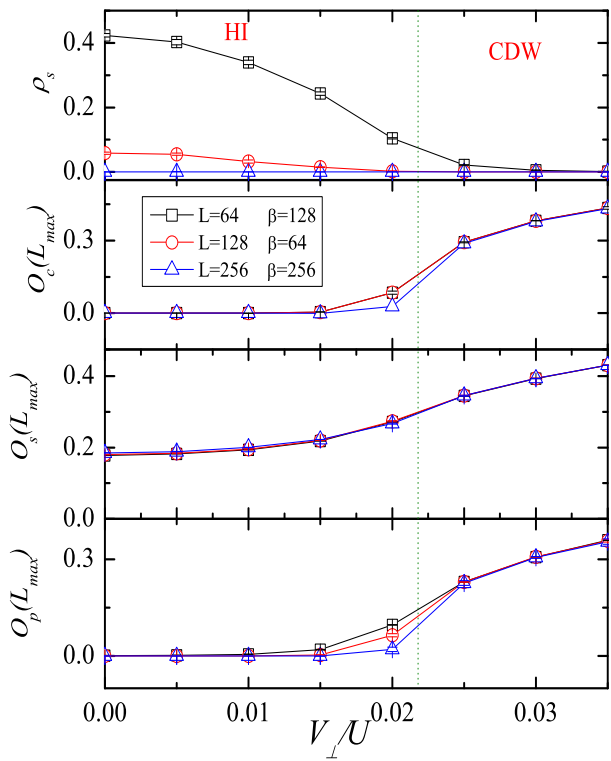}
\caption{\label{t_0.25} (Color online) Quantities $\rho_s$, $O_c(L_{max})$, $O_s(L_{max})$ and $O_p(L_{max})$ versus $V_{\bot}/U$ at $V_{\parallel}/U=3/4$ and $t/U=1/4$.}
\end{figure}

First, we explore the stability of HI against interchain interaction.
Figure~\ref{t_0.25} shows the Monte Carlo results for $t/U=1/4$ to reveal whether and when the HI disappears.
For each considered values of $V_{\bot}/U$, we find in the thermodynamic limit that $\rho_s=0$. Therefore the system is
always insulating.  As $V_{\bot}/U$ increases, a phase transition occurs at $V_{\bot}/U \approx 0.022$ beyond which the bosons form a CDW phase. This transition is reflected by a sharp increase of $O_c(L_{max})$ from $0$  to
a finite value. Furthermore, the Haldane string order is robust under weak interchain interactions and persists for $0\leq V_{\bot}/U<0.022(4)$. Therefore, the HI phase is stable under weak interchain interactions.

\begin{figure}
\centering
\includegraphics[angle=0,width=8cm,height=8cm]{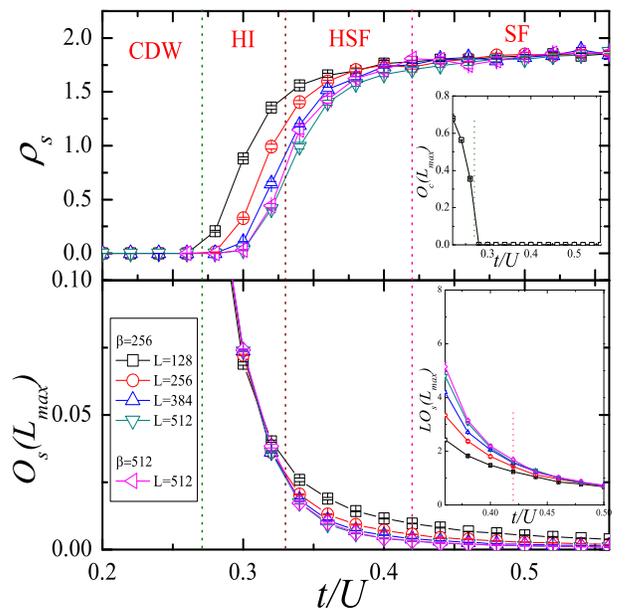}
\caption{\label{V2_0.04}  (Color online) Quantities $\rho_s$, $O_s(L_{max})$ and  $O_c(L_{max})$ versus $t/U$ at $V_{\parallel}/U=3/4$ and $V_{\bot}/U=1/25$. The inset of lower panel is for scaled string order parameter $L O_s(L_{max})$ and focuses on HSF-SF transition.}
\end{figure}

As interchain repulsions are present, the topological HSF phase emerges. As shown in the phase diagram (Fig.~\ref{pd_v1_0.75}), the HSF phase is sandwiched between HI and SF phases.
It is crucial to analyze the effects induced by strengthening the hopping of the bosons which are in a HI phase, and to see how they become SF. As demonstrated in Fig.~\ref{V2_0.04} for $V_{\bot}/U=1/25$,
we determine four quantum ordered phases: as $t$ increases, the CDW, HI, HSF and SF phases appear sequently, and are separated by three transitions.
The CDW-HI phase transition occurs at $t/U=0.27(1)$. While $O_s(L_{max})$ is robust in both sides of the transition point, the crystalline order parameter $O_c(L_{max})$ vanishes in HI phase and remains robust in CDW phase.
 The transition point can be determined via the sudden decrease of $O_c(L_{max})$.
The superfluidity emerges at $t/U\approx0.33$ which is estimated via the evaluation on where the variation of $\rho_s$ is most drastic. Surprisingly, the Haldane string order does not vanish immediately.
The existence of both string order and superfluidity reveals a novel topological
phase which we term as HSF. As $t/U$ further increases, the string order finally vanishes while SF density persists, and the system becomes SF.
The precise determination of HSF-SF transition point is difficult since the amplitude of string correlation in HSF phase is
already weak (but finite). Here we employ a similar treatment as that in Ref.~\cite{Batrouni2014},
and plot $LO_s(L_{max})$ for different lattice sizes. In the SF side and as $L \rightarrow \infty$, one expect $O_s(L_{max}) \propto L^{-1}$. Therefore, the $LO_s(L_{max})$ curves of different sizes become to merge together at transition point, which is estimated to be $t/U=0.42(1)$. Correspondingly, a more or less useful signal can be found from $\rho_s$ data: in the SF side $\rho_s$ is almost a constant as $t/U$ changes while in the HSF side it varies more quickly.
These evidences together indicate a HSF-SF transition, occurring at $t/U=0.42(1)$.

\begin{figure}
\centering
\includegraphics[angle=0,width=8cm,height=9cm]{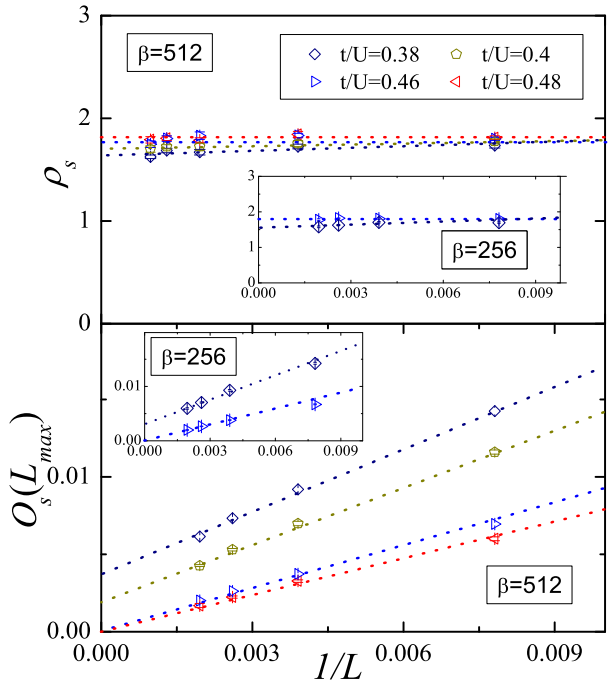}
\caption{\label{scalings} (Color online)  Finite-size scalings of $\rho_s$ and $O_s(L_{max})$ at different $t/U$, with $V_{\parallel}/U=3/4$ and $V_{\bot}/U=1/25$. The main (inset) panels are for $\beta=512$ ($256$). The $t/U$ ratios are $0.38$ (HSF), $0.4$ (HSF), $0.46$ (SF) and $0.48$ (SF). Dotted lines indicate extrapolations to the infinite chain.}
\end{figure}

To further confirm the existence of the HSF phase and the HSF-SF transition, we perform finite-size scaling on $\rho_s$ and $O_s(L_{max})$ at low temperatures, for different $t/U$ that are not far away from the estimated HSF-SF transition point. As illustrated in Fig.~\ref{scalings} for both HSF and SF phases, we confirm a finite SF density as $L \rightarrow \infty$. When $L$ is large enough, a good linear fit between $O_s(L_{max})$ and $1/L$ is achieved in the SF phase. In HSF phase, however, $O_s(L_{max})$ extrapolates to a finite value as $L \rightarrow \infty$.
In all these scalings the temperature effects are negligible, indicating that ground-state information is already obtained. In short, the existence of HSF and SF phases are confirmed respectively in different sides of $t/U = 0.42(1)$, which is therefore a reliable estimate of transition point.

\begin{figure}
\includegraphics[angle=0,width=8.5cm,height=8cm]{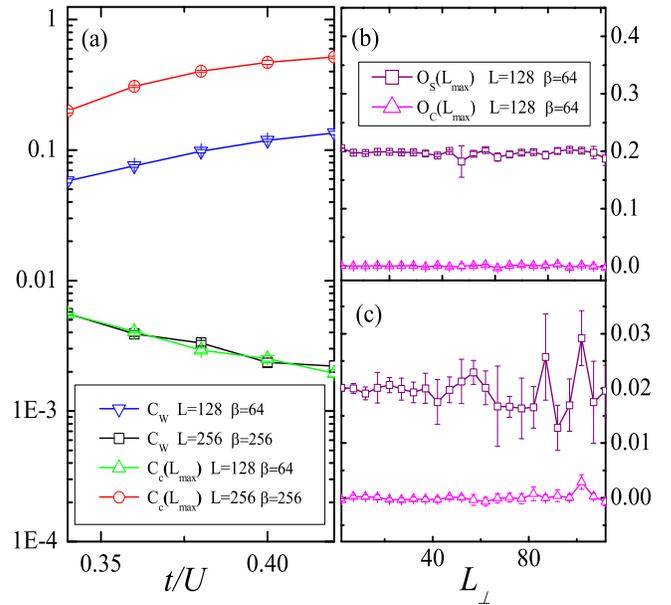}
\caption{(a) Correlations $C_W$ and $C_c(L_{max})$ in HSF region with
$V_{\parallel}/U=3/4$ and ${V_\bot}/U=1/25$. Quantities $O_s(L_{max})$ and $O_c(L_{max})$ versus chain number in (b) HI at $V_{\parallel}/U=3/4$, $V_{\bot}/U=0.01$ and $t/U=1/4$, and (c) HSF at $V_{\parallel}/U=3/4$, ${V_\bot}/U=1/25$ and $t/U=0.36$.}
\label{merged}
\end{figure}

We then discuss the underlying picture of HSF phase. To address the possibility of phase separation,
we collect equal-time snapshots of worldline configurations in deep HSF phases, using
different initial states and random number seeds in the simulations. In none of the snapshots we find phase
separation which can be signaled by separated regions of Haldane string order.
One can imagine two scenarios on interchain correlations: (i), the symmetry between
the chains is broken and the bosons on one chain become SF and on the other form HI;
(ii), superfluidity and Haldane string order are both present in either chain.
To distinguish between these possibilities and get an in-depth understanding of HSF,
we calculate two types of interchain correlations:
\begin{eqnarray}
&C_{\rm W}&=\langle (W_{\rm T}W_{\rm B})^2\rangle, \nonumber \\
&C_{\rm c}(|i-j|)&=\langle \delta n_{{\rm T},i} \delta n_{{\rm T},j} \delta n_{{\rm B},i} \delta n_{{\rm B},j}\rangle.
\end{eqnarray}
Where we set $|i-j|=L_{max}=L/2$; the symbols {\rm T} and {\rm B} indicate that the measurements are taken on top and bottom chains, respectively. As shown in Fig.~\ref{merged}(a) for a HSF regime that $C_{\rm W}>0$ and $C_{\rm c}(L_{max})>0$, which hold as $L$ increases. Thus, in the HSF phase the interchain correlations stabilize both non-diagonal and diagonal orders. These findings indicate that both chains exhibit a robust superfluidity as well as a finite non-local Haldane order. Scenario (ii) is therefore validated and what we find on each chain is a stable HSF phase.

\section{Summary and discussions}~\label{s5}
We have mapped out the significant part of the ground-state phase diagram of a weakly coupled chains of extended Bose-Hubbard model. In particular, we identified a novel topological phase, the HSF phase, which has both non-local Haldane order and superfluidity. %A natural extension is to multi-chain versions of Hamiltonian~(\ref{hamiltonian}).
As shown in Figs.~\ref{merged}(b) and (c), we demonstrate how the non-local Haldane order and the CDW order vary as the chain number increases. It is found that the non-local orders in HI and HSF phases are both stable and nearly unchanged within the uncertainties indicated by error bars.

Our proposal serves as a rather controllable approach to realize the TSF, without any involvement of spin-orbit coupling and Zeeman field.
The existence of the HSF is directly testable by cold bosons in a two-leg optical ladder, where boson-boson interactions can be tuned via Feshbach resonances.
Further, as indicated in Ref.~\cite{Atala2014}, the non-local characteristics of cold bosons can be measured on either leg of an optical ladder. The other realization of the present Hamiltonian may be the bosonic mixture on an optical lattice, which is achievable within state-of-the-art experimental capabilities~\cite{Catani2008,Thalhammer2008}.

Finally, we wish to remark that there are rare examples of coexisting diagonal and off-diagonal orders in many-body systems.
A paradigm is the SS phase on a lattice, which hosts both crystalline order and superfluidity,
mostly induced by {\it simultaneously} broken translational and gauge symmetries, respectively~\cite{Boninsegni2012}.
The HSF phase, a quantum phase that has both non-local Haldane string order and superfluidity, is a new paradigm.

%\textit{Acknowledgments.-}

\begin{acknowledgments}
We acknowledge Yuxin Zhao, Xiaosen Yang and Huaimin Guo for useful discussions.
This work was supported by the GRF (Grant Nos.
 HKU173051/14P and HKU173055/15P), the CRF (HKU8/11G) of Hong Kong.
One of us (JPL) was supported by the National Natural Science Foundation of
China (Grant No. 11405003) and
the Anhui Provincial Natural Science Foundation (Grant No. 1408085MA15).
\end{acknowledgments}

\end{document}